# Approaches to modelling irradiation-induced processes in transmission electron microscopy

Stephen T. Skowron,[a] Irina Lebedeva,[b,c] Andrey Popov,[d] Elena Bichoutskaia[a,*]

The recent progress in high-resolution transmission electron microscopy (HRTEM) has given rise to the possibility of *in situ* observations of nanostructure transformations and chemical reactions induced by electron irradiation. In this article we briefly summarise experimental observations and discuss in detail atomistic modelling of irradiation-induced processes in HRTEM, as well as mechanisms of such processes recognised due to modelling. Accurate molecular dynamics (MD) techniques based on first principles or tight-binding models are employed in the analysis of single irradiation-induced events, and classical MD simulations are combined with a kinetic Monte Carlo algorithm to simulate continuous irradiation of nanomaterials. It has been shown that sulphur-terminated graphene nanoribbons are formed inside carbon nanotubes as a result of an irradiation-selective chemical reaction. The process of fullerene formation in HRTEM during continuous electron irradiation of a small graphene flake has been simulated, and mechanisms driving this transformation analysed.

## 1. Experimental imaging of electron irradiation-induced processes

Recent advances in electron microscopy, in particular the implementation of aberration corrections of electromagnetic lenses, have led not only to the possibility of imaging light atoms but also to the observation of single atom dynamics.[1,2,3] These developments potentially provide a tool for direct measurements of characteristics of dynamic processes, including diffusion coefficients, cross-sections and chemical constants. The progress in experimental high-resolution transmission electron microscopy (HRTEM) techniques has also resulted in the possibility of studying atomic scale structure transformation *in situ*, ultimately leading to the imaging of a single irradiation-induced chemical reaction.

The electron irradiation-induced processes in HRTEM have attracted considerable interest in recent decades; these include the creation of single vacancies and other atomic scale defects,[1,2,4–9] the transformation of carbon[7-26] and boron nitride[27–30] nanostructures, leading to the formation of entirely new nano-objects,[12,17–26] chemical reactions[25,26,31–36] and irradiation-activated molecular motion.[37,38,39] The creation of atomic scale lattice defects has been extensively studied in graphene[1,2,4,5] and carbon nanotubes.[6] The formation of vacancies arising from electron beam (e-beam) knock-on damage[1,4] and the removal of these vacancies via a chemical reaction with trapped adsorbates[1] was observed *in situ* for a graphene layer, as well as the formation and subsequent relaxation of isolated Stone-Wales (SW) defects and multiple 5-7 defects.[2,5] It was also found that in addition to vacancies, nanometre-scale holes are repaired by mobile carbon ad-atoms emitted from neighbour graphene layers[12] or hydrocarbon contamination.[13] Both amorphous and crystalline patches of healed graphene were observed, with crystallisation dominating at increased temperatures.[12] The focused e-beam of an aberration-corrected scanning transmission electron microscope was also used to create individual vacancies in single-walled nanotubes, and the subsequent reconstruction of these vacancies was observed.[6] The migration and reconstruction of divacancies,[5,7] and the fluctuation in boundary shape between domains of different orientation[8] were induced by the HRTEM e-beam on a graphene layer. Irradiation damage in nitrogen-doped graphene was found to initiate at lower electron energy via the removal of a nitrogen-neighbouring carbon atom.[9]

The further transformation of graphene structure containing defects depends on the experimental conditions such as temperature, energy and density of the e-beam. The atomic scale vacancies have been shown[10–12] to evolve into holes 3-10 Å in diameter upon e-beam irradiation. Other studies of irradiation-induced transformation of vacancies in graphene reveal that divacancies also merge into extended multi-vacancy structures.[7] The detailed analysis of HRTEM images shows that linear configurations of divacancies can be formed involving four-atom rings. The e-beam driven rotation of bonds subsequently results in a configuration consisting of a core of hexagons surrounded by a chain of alternating pentagons and heptagons. This presents an interesting route to making a new two-dimensional material, an amorphous carbon membrane with a random arrangement of polygons.[7]

A large number of HRTEM studies have been devoted to irradiation-induced evolution and reconstruction of two-dimensional crystal edges.[11,12,14–17] It was concluded that at room temperature the zigzag edge of graphene is more stable under electron irradiation than the armchair edge,[11,14] whereas at 900 K the armchair edge becomes more stable.[12] The formation of dangling bonds and multi-member rings near the edge,[15,17] as well as the stepwise migration of atoms along the edge[11,14] have been shown to contribute to its reconstruction. The supply of carbon atoms by adsorbates can lead to repair of the edge.[12,14] The probability of edge reconstruction is found to increase at higher temperatures.[12] The transition from a zigzag edge, terminated by hexagons, to the reconstructed edge, terminated by alternating pentagons and heptagons, has been demonstrated.[17] It was also shown that a combination of Joule heating and electron irradiation leads to layer-by-layer peeling of multi-layer graphene, compared to the perforated structure formation under electron irradiation without Joule heating.[16]

Unlike graphene, the knock-on damage of hexagonal boron nitride (*h*-BN) is more straightforward.[27–30] Although vacancies can be formed in *h*-BN, topological defects have not been observed. Triangular[27–30] and hexagonal holes[28] with edges terminated by nitrogen atoms have been seen to originate from the vacancies formed by the removal of the lighter boron atoms. The annihilation of vacancies via recombination with ad-atoms produced by electron impacts was also observed, analogous to graphene.[29]

The creation of irradiation-induced defects in a nanostructure can lead to a considerable realignment of bonds, such that a transformation into an entirely new species takes place.[12,14–26] One of the most striking examples of such a transformation is the formation of a fullerene from an initially flat graphene flake.[18] Others include the transformation of polyhedral graphitic nanoparticles into quasi-spherical onions,[40,41] the formation of double-walled nanotubes from single-walled nanotubes filled with fullerenes,[42] and the production of a nanotube-encapsulated fullerene from carbon clusters, partially built from regions of defects in the nanotube wall.[19] The transformations of amorphous carbon on a graphene surface into a graphene layer rotated with respect to the supporting graphene layer,[20] and of a two-layer graphene nanoribbon into a flat nanotube[12] have been reported. New nano-objects can be engineered by the e-beam via the knock-on removal of a considerable portion of atoms from the initial structure. Flattened single-walled carbon nanotubes have been produced in this way directly from bilayer graphene.[43] In this experiment, a TEM in scanning mode has been first used to produce bilayer graphene nanoribbons with predefined width, which during subsequent observation in HRTEM underwent transformation into a flattened carbon nanotube. Single, double and even triple monoatomic carbon chains have also been produced in HRTEM from graphene ribbons.[12,17,21]

Interesting new species were obtained following irradiation-induced transformations of organic molecules in the confined space of nanotubes. These transformations strongly depend on the chemical elements present in precursor molecules. For example, a trilobate structure was formed from three coalescing $La@C_{82}$ endofullerenes.[23] Carbon atoms generally tend to form a graphene-like structure within the nanotube cavity. If precursor molecules contain no elements able to saturate the edge $sp^2$-carbons, a new inner wall forms inside the carbon nanotube. The new wall can be empty, for example if it forms from fullerenes. It can be also filled with a nanocrystal, as in the case of the transformation of $Pr_2@C_{72}$ endofullerenes into an additional inner wall of a multi-walled nanotube confining a $PrC_2$ nanocrystal.[24] If atoms that can saturate dangling bonds at the graphene-like edge are present in the mixture, a graphene nanoribbon forms. Sulphur-terminated graphene nanoribbons have been obtained in this way through the transformation of fullerenes with attached sulphur-containing groups[25] and a sulphur-containing molecule, tetrathiafulvalene,[26] inside carbon nanotubes.

The irradiation-induced activation of chemical reactions between metals and carbon nanostructures has been studied in experiments including rather straightforward cases of metal-graphene interactions[13,31–34] and the complicated processes of transformation in which metal clusters or carbon-metal nanostructures are located inside carbon nanotubes.[35–37] Multiple observations have been made of graphene-metal interactions activated by electron irradiation, including the collective motion of the Au atom and nearby carbon atoms,[31] hole formation in graphene assisted by Pd,[13,32,33] Ni, Ti, Al,[32,33] and Cr[33] atoms and clusters, and an enhanced rate of hole formation in graphene near iron clusters[34]. In the presence of catalytically active atoms of Re inserted into nanotubes, nanometre-sized hollow protrusions on the nanotube sidewall are formed, thus confirming that the nanotube sidewall can be engaged in chemical reactions from the inside.[44] The complex process of the transformation of a single-walled carbon nanotube with an Os cluster inside has been reported.[35] In this case carbon bonds, weakened by the interaction with the metal cluster, break upon the electron irradiation, the newly formed edges rearrange into closed caps, metal atoms rearrange to form parallel 6-7 atom long chains between the caps, and the contact between the cluster and the nanotube is broken. Another multiple stage irradiation-induced transformation has been observed for $Dy@C_{82}$[36] and $La@C_{82}$[37] endofullerenes inside a single-walled nanotube. It consists of dimerization of endofullerenes with the release of Dy atoms (or La atoms), transformation of the newly formed dimers into an inner wall and aggregation of Dy atoms into a cluster (or La atoms into $La_2C$ nanocrystal), and rupture of the nanotube cap assisted by the Dy cluster (or escape of $La_2C$ from the interior of the carbon nanotube). A very interesting electron irradiation-induced process observed in HRTEM is the activation of molecular motion.[37,39] The e-beam activated motion of a $La_2C$ nanocrystal-nanotube system within a host nanotube[37] and the passage of a hydrocarbon chain through a hole in a single-walled carbon nanotube have been observed.[39] Additional aspects of nanostructure transformation under irradiation influence are further discussed in reviews.[45,46]

The above advances in experimental imaging techniques require theoretical solutions capable of treating the dynamic evolution of structures under the e-beam. In this article the state-of-the-art and recent progress in atomistic modelling of electron irradiation-induced processes is discussed.

## 2. Atomistic modelling of electron irradiation-induced processes

### 2.1 Computational approaches to modelling electron irradiation-induced processes

For the purpose of the present article, electron irradiation-induced processes are divided into two principal categories, and referred to as *simple* and *complex* irradiation-induced process. Simple irradiation-induced processes occur as a result of a single transformation event caused by an electron impact with atoms (for example the formation of vacancies or SW defects in a pristine graphene or nanotube structure) or as a result of a few predetermined transformation events (for example the transformation and migration of vacancies and other defects in graphene and the irradiation damage of *h*-BN). The complex irradiation-induced processes occur in several stages such that numerous new types of unpredicted local atomic structures emerge, which do not form in the absence of the e-beam. Their subsequent transformations should be taken into account when considering the entire e-beam assisted transformation process.

The studies of simple and complex irradiation-induced processes by computer simulation require different modelling techniques. Predetermined transformation events of a simple process can be computed in a rather small modelling cell containing tens of atoms, with a total simulation time of about 100 ps, sufficient to describe a single transformation event. Simple irradiation-induced processes are therefore studied with the use of molecular dynamics (MD) techniques based on first principles or tight-binding models. The simulation of complex processes requires hundreds of atoms in the modelling cell and a total simulation time of the order of tens of nanoseconds necessary to include hundreds of subsequent transformation events. Such complex processes are studied using MD with semi-empirical force fields in order to reduce computational costs.

There follows a consideration of some methodological aspects of these approaches and an overview of the main results described in the literature. New striking results utilising both approaches are also presented: first principles MD simulations showing that sulphur-terminated graphene nanoribbons inside carbon nanotubes can form as a result of irradiation-selective chemical reactions, and classical MD simulations of the graphene-fullerene transformation.

**2.2 Modelling of single irradiation-induced events**

The simple irradiation-induced processes described in section 2.1, in contrast to the complex processes affected by subsequent transformations, have been studied in detail for graphene,[9,48,51,54] carbon and boron nitride nanotubes,[9,47-51,55] and $h$-BN.[52] As explained in section 1, in $h$-BN electron irradiation-induced reactions other than a simple knock-on emission of atoms have not been observed, making this material well suited to the modelling of single events. However, carbon nanomaterials are generally subjected to more complex processes, thus requiring more advanced modelling methods.

Simple irradiation-induced processes due to a single electron impact are commonly studied using *ab initio* MD (AIMD) or density functional theory-based tight binding (DFTB) methods. These methods are used to calculate the displacement (or emission) threshold energy, i.e. the minimum electron kinetic energy required for an atom to permanently leave its lattice position. For anisotropic systems the displacement threshold energy depends on the emission direction; in the case of nanotubes it is dependent on the angular position of the atom around the tube circumference.[47] In TEM, atom emission occurs when the threshold energy is lower than the maximum energy transferred from an electron to the atom, which corresponds to an atom emission angle of 0°, i.e., when the atom is emitted in the same direction as the e-beam. The maximum transferred energy is dependent on the accelerating voltage used.

The calculations of the displacement threshold energy revealed multiple novel features of irradiation-induced atom emission in a variety of nanomaterials. Irradiation damage in both carbon and boron nitride nanotubes was shown to occur at two separate regimes;[47] at low energies emission primarily occurs from the upper and lower parts of carbon nanotubes and involving boron in BN nanotubes, while at higher energies the sides of carbon nanotubes were preferentially damaged and nitrogen emission was more favourable for BN nanotubes. By calculating the displacement threshold energy as a function of mechanical strain of carbon nanotubes, it was discovered that certain nanotubes under a small amount of strain had a surprising increase in stability under irradiation.[48] The increased susceptibility of nitrogen atoms to irradiation damage was found to cause the lowered stability of nitrogen-doped carbon nanotubes compared to pristine nanotubes.[49] The effect of carbon nanotube diameter and chirality on electron irradiation damage was studied,[50] showing the preferred formation of divacancies compared to monovacancies. Carbon nanotubes with smaller diameters were found to have smaller displacement threshold energies, providing an answer to experimental observation that the inner walls of multi-walled carbon nanotubes are preferentially damaged by electron irradiation.

In addition to atom emission, the same approach has been applied to the study of SW-type transformations in pristine graphene and graphene containing divacancy structures.[51] This study showed the presence of two stages in SW transformations responsible for the inter-conversion of divacancy structures and divacancy migration. It also showed that the central atoms in reconstructed divacancies have larger displacement threshold energies than those of pristine graphene, providing a possible explanation for the observation that under continuous electron irradiation defects tend to grow into large amorphous structures rather than collapsing into holes.

The AIMD simulations were used to determine the displacement threshold energies in $h$-BN,[52] giving significantly different results to those produced by DFTB. The discrepancy between the two methods was ascribed to the inadequate description of the charge transfer between boron and nitrogen in DFTB, while results for the removal of carbon in a graphene sheet were comparable for both methods. These simulations were extended to macroscopic time scales through the use of kinetic Monte Carlo simulations with displacement rates based on the AIMD calculations; this allowed a direct comparison between experiment and theory to be made. Note that irradiation damage of $h$-BN can be reliably described using this simple model, as the atom displacements are the only irradiation-induced events observed for $h$-BN.

The AIMD study of electron irradiation on graphite[53] revealed the effect of the azimuthal and electron scattering angles on the nature of the produced defects; SW and Frenkel pair defects were among the typical irradiation-induced defects formed. The experimental observations, discussed in section 1, showing that electron irradiation damage of nitrogen-doped graphene begins with the removal of the nitrogen-neighbouring carbon atom were confirmed with AIMD simulations of the displacement threshold energies.[9] The AIMD method was also used to study the stability of atoms at the edges of graphene nanoribbons of various widths and edge configurations.[54] The thermodynamic stability of graphene edges was determined to differ from the electron irradiation stability, with armchair edges showing a surprising degree of stability under irradiation. This difference in the thermodynamic and electron irradiation stabilities indicates that chemical bonds that do not correspond to the thermodynamic equilibrium in the system can survive under electron irradiation. This conclusion leads to the idea of selective control of chemical reactions by electron irradiation in heterogeneous systems consisting of atoms of several elements. The possibility of irradiation-selective chemical reactions is considered in Section 4.1.

### 2.3 Modelling of continuous irradiation-induced events

As previously mentioned, a direct simulation of continuous irradiation-induced events is required to understand the mechanisms of complex processes in which new unpredicted local atomic structures form. There exist a number of such simulations using MD techniques with semi-empirical force fields, which include the irradiation damage of carbon nanotubes[55] and graphene nanoribbon edges,[15] the cutting of graphene nanoribbons by the e-beam,[56] the formation of holes[3] and flattened single-walled carbon nanotubes[43] in bilayer graphene, and the first stages of irradiation damage of multilayer graphene and single- and double-walled nanotubes on substrates.[57]

MD simulations confirmed that irradiation damage of carbon nanotubes increases with reduction in the nanotube diameter.[55] These simulations also show that the rate of atom removal by electron impact halves as the temperature is increased from 300 to 1000 K. This leads to the formation of an amorphous structure with a smaller diameter at higher temperatures, while at lower temperatures the nanotube begins to separate. MD simulations of the irradiation damage of graphene nanoribbons with different edges show that at 300 K for an incident electron energy of 60 keV the zigzag edge is dynamically more stable than the armchair edge.[15] At 300 K a superfluous number of carbon chains and rings with an extended length of up to 20 atoms is formed, however these chains and rings are absent if the simulation is performed at a temperature of 2000 K. The elevated simulation temperature also leads to an increased number of SW defects created by irradiation in the nanoribbon structure. MD simulations of cutting graphene nanoribbons with the e-beam, performed at two different temperatures, show the formation of different structures during the irradiation damage.[56] Namely, amorphous structure and the shrinkage of the width of the nanoribbon with the subsequent formation of long carbon monoatomic chains was only seen at 1500 K, whereas cutting with a relatively smooth edge was observed at 300 K.

Tubular structures and nanotube junctions have been obtained in MD simulations by the creation of adjacent holes arising from electron irradiation in bilayer graphene.[56] These new nano-objects have been found in the case where the irradiation damage occurs at a temperature of 300 K with the subsequent annealing of the system occurring after irradiation at 1500 K. MD simulations devoted to the influence of different substrates on the irradiation-induced damage of carbon nanostructures reveal that the increase of defect formation in multilayer graphene and single- and double-walled nanotubes on substrates is due to backscattered electrons.[57]

The MD simulations described above indicate an essential role of temperature in irradiation-induced processes. However, these simulations do not allow for the discrimination between the role of temperature in a collision between an electron and an atom nucleus, and thermally induced processes taking place simultaneously with irradiation-induced processes. A typical electron flux on a sample in HRTEM is of the order of $10^7$ electrons per $nm^2$ per second, and there are approximately 40 carbon atoms per 1 $nm^2$ of graphene network. Assuming that the atom ejection cross-section is of the order of 10 barn ($10^{-9}$ $nm^2$) there is an interval of 2.5 s between the subsequent irradiation-induced events taking place in the area of 1 $nm^2$. Thus there are two characteristic features of irradiation-induced processes in HRTEM: 1) a very long time period between irradiation-induced events during which the annealing of the formed defects and even self-healing of the nanostructure can occur, 2) the majority of interactions between incident electrons and atoms do not lead to significant structural changes. For example, MD calculations show that vacancies in graphene nanoribbons less than 2 nm in width move to the edge within few ms.[58] This time is several orders of magnitude shorter than the time period between the subsequent irradiation-induced events which provide effective self-healing of the nanoribbon. Thus structural relaxations between irradiation-induced events should be taken into account in atomistic simulations of irradiation-induced processes. Clearly the large time between irradiation-induced events in HRTEM is not directly accessible by MD techniques. Kinetic Monte Carlo-type algorithms are therefore required, which allow for an adequate and effective description of the transformation of nanostructure under the e-beam, and account for the thermally induced structural changes between irradiation-induced events.

Due to the stochastic nature of non-equilibrium processes in systems containing hundreds of atoms, these processes could evolve in different ways despite identical starting conditions. For example, MD simulations show that both empty fullerenes and endofullerenes with a nickel core can be formed during the Ni-assisted transformation of graphene flakes,[59] and that different dynamical behaviour is possible for the retraction of telescopically extended graphene flakes on graphene surface.[60] The possibility of different transformation paths should be therefore taken into consideration. Several simulation runs with identical starting geometry and conditions and a fresh random seed are required to ensure that the process occurs in a single unique way, or even hundreds simulation runs to obtain good statistics of the transformation process. In section 4.2, it is shown that upon electron irradiation damage of a small graphene flake, both evaporation of the flake and formation of a fullerene are possible. The stochastic nature of irradiation-induced processes represents an additional problem, which increases computational time required for atomistic simulations of the structure transformations.

### 2.4 *CompuTEM* algorithm

A simple model describing a uniform flux of incident electrons interacting with the nanostructure kept at constant temperature or energy cannot be used effectively for the adequate simulation of complex irradiation-induced processes observed in HRTEM. We propose an effective algorithm for simulation of such processes, in which only interactions between incident electrons and atoms leading to changes in the atomic structure are considered, and in which irradiation-induced events and annealing of the structure between these events are computed independently at different temperatures.[61]

As illustrated in Fig. 1, a random irradiation-induced event occurring in a sequence of such events is described as follows: 1) the nanostructure is equilibrated at a temperature corresponding to experimental conditions in HRTEM, 2) each atom in the nanostructure is classified with respect to the number and strength of its chemical bonds, 3) the probability of an irradiation-induced event (such as atom removal and/or changes to the local atomic structure) is assigned to each atom in accordance with the atom type determined at step 2, so that the sum of the probabilities over all atoms and all considered types of events for each atom is equal to unity, 4) a single random irradiation-induced event is introduced, 5) MD simulation at a temperature corresponding to experimental conditions for a duration sufficient for bond reorganisation or atom removal, 6) MD simulation at the elevated temperature taking into account the structure relaxation after the irradiation-induced event.

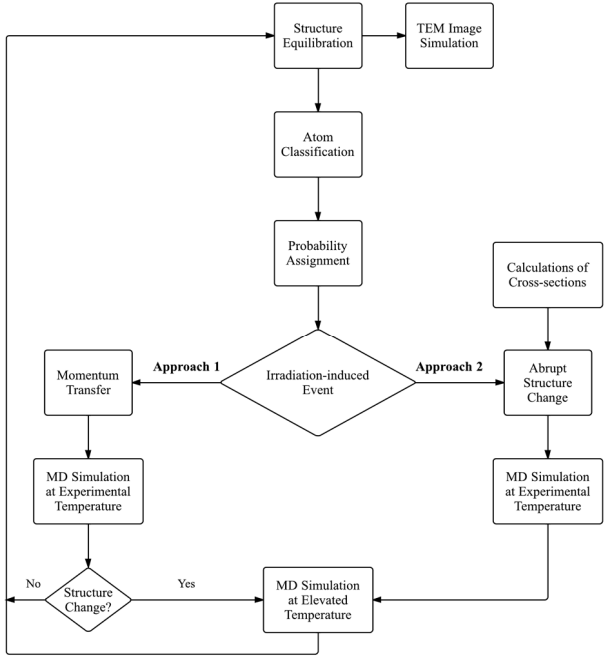

**Fig. 1.** The CompuTEM algorithm, with two approaches for describing the interaction between beam electrons and sample atoms. Approach 1: a full transfer of momentum from electron to atom is modelled for each impact event. Approach 2: successful impacts are modelled as abrupt changes in atomic structure, corresponding to the outcome of selected processes, weighted by the cross-sections of these processes.

A software implementation of this algorithm, *CompuTEM*,[61] also provides a sequence of images of sample evolution with time during its observation in HRTEM for given experimental conditions. In *CompuTEM*, two main computational parts, molecular dynamics and image simulations, are linked by the experimental value of the electron flux, which allows the up-scaling of MD simulation time to the experimental time and determines the signal-to-noise ratio of the simulated images. Within the general scheme described above, we consider two approaches with different realisations of step 4. In approach 1 momentum is transferred to a given atom from the incident electron, as the time of electron-atom interaction in HRTEM is considerably less than the time of a single MD simulation step. In approach 2 the irradiation-induced event (for example atom removal or SW transformation) is introduced into the system as a corresponding abrupt change of the local atomic structure.[61]

Evidently, probabilities of random irradiation-induced events are proportional to the cross-sections of these events for the interaction between the incident electron and nucleus, which implies that prior knowledge of these cross-sections is required. In approach 2 the cross-sections are calculated for all considered irradiation-induced events using AIMD or DFTB techniques or obtained from experiment. In approach 1 the minimum threshold energy for every irradiation-induced event needs to be calculated for all different types of atoms using the same modelling conditions (such as force field or temperature) as used for simulating the irradiation-induced process. The minimum threshold energy determines the total scattering cross-section, thus including all possible irradiation-induced events for the given type of atom.

The total structure evolution time at experimental conditions can be expressed as a sum of the time periods between subsequent irradiation-induced events. The time period between the events, $t_{ev}$, is defined as the inverse of the product of the overall cross-section corresponding to all considered irradiation-induced events, $\sigma$, and the electron current density, $j$

$$t_{ev} = 1/j\sigma \qquad (1)$$

In approach 1, the overall cross-section is assigned to each type $u$ of atoms as $\sigma = \sum_u N_u \sigma_u$, where $\sigma_u$ is the cross-section corresponding to all considered irradiation-induced events; $N_u$ is the number of atoms of $u$-th type in the nanostructure. In approach 2, the overall cross-section is defined as $\sigma = \sum_u N_u \sum_v \sigma_{uv}$, where $\sigma_{uv}$ is the cross-section for irradiation-induced event $v$ for atoms of $u$-th type. Equation (1) gives the rate of structure evolution under the influence of the e-beam and allows a direct comparison of the simulated process with experimentally observed dynamics.

If the kinetic energy transferred to an atom from an incident electron is comparable to the value of the threshold energy then changes in the local atomic structure are mainly determined by thermal deformation of the structure and by the velocity of the atom at the moment of collision. In approach 1, if the transferred energy is sufficient to induce changes in the local atomic structure such an interaction is referred to as a *successful impact*. However, *all impacts*, including unsuccessful, should be included in the estimation of the rate of structure evolution using equation (1). If the irradiation-induced process is simulated using approach 1, the change in the local atomic structure needs to be confirmed after step 5 by an analysis of the impacted atom topology. If there is no change, then stage 6 should be omitted. Note that the largest part of the computational time corresponds to stage 6. The simulation of the entire irradiation-induced process, accounting for unsuccessful impacts, does not lead to a considerable increase in the computational time if the proportion of successful impacts is about 0.1 or greater.

Irradiation-induced and thermally induced processes of nanostructure transformation often occur in very similar ways. For example, double-walled nanotubes can be formed from single-walled nanotubes filled with fullerenes both under electron irradiation[42] and at high temperature.[62] However, irradiation-induced processes in TEM usually take place at a temperature at which analogous thermally induced processes are not possible. The elevated temperature $T_{rel}$ should be therefore chosen such that the reconstruction of irradiation-induced structure changes is accounted for, but thermally induced transformations of the pristine structure are excluded. In other words, the following condition should be fulfilled

$$N_{ir} t_{rel} \ll t_{th} \quad (2)$$

where $N_{ir}$ is the number of irradiation-induced events during the simulation of the irradiation-induced process, $t_{rel}$ is the relaxation time between irradiation-induced events used at step 6 of the algorithm, and $t_{th}$ is the characteristic simulation time required for the thermally induced process analogous to the considered irradiation-induced process to take place at the elevated temperature $T_{rel}$. The dependence of the characteristic time $t_{th}$ on temperature is determined by Arrhenius equation for the majority of thermally induced processes

$$t_{th} = \tau_0 \exp(E_a / k_B T) \quad (3)$$

where $\tau_0$ is the pre-exponential factor, $E_a$ is the maximum activation energy along the path of the process, and $k_B$ is the Boltzmann constant. This dependence is particularly shown in classical MD simulations of the thermally induced graphene-fullerene transformation.[59] The Arrhenius equation allows an estimation of the characteristic time $t_{th}$ at the temperature $T_{rel}$ to be made via simulations of the process without electron irradiation at temperatures $T \gg T_{rel}$. At these temperatures the timescale of the thermally induced process is tractable with classical MD techniques.

In section 4.2 approaches 1 and 2 are applied to the simulation of a graphene-fullerene transformation. The advantages of both approaches and possible areas of application for the simulation of different irradiation-induced process are discussed in the conclusion.

## 3. Methods

### 3.1 Modelling of electron collisions

In approaches 1 and 2, the electron collisions are considered in the framework of the standard theory of elastic electron scattering between a relativistic electron and a nucleus.[63,64] The differential scattering cross-section is calculated according to the McKinley and Feshbach formula[65]

$$\sigma(\theta) = \sigma_R \left[ 1 - \beta^2 \sin^2 \frac{\theta}{2} + \pi \frac{Z e^2}{\hbar c} \beta \sin \frac{\theta}{2} \left( 1 - \sin \frac{\theta}{2} \right) \right], \quad (4)$$

where $Z$ is the nuclear charge, $\beta = V_e / c$ is the ratio of electron velocity $V_e$ to the speed of light $c$, $\hbar$ is the Planck constant, $e$ is the electron charge, $\theta$ is the electron scattering angle (Fig. 2), $\sigma_R$ is the classical Rutherford scattering cross-section

$$\sigma_R = \left( \frac{Z e^2}{2 m_e c^2} \right)^2 \frac{1 - \beta^2}{\beta^4} \csc^4 \left( \frac{\theta}{2} \right), \quad (5)$$

and $m_e$ is the electron mass.

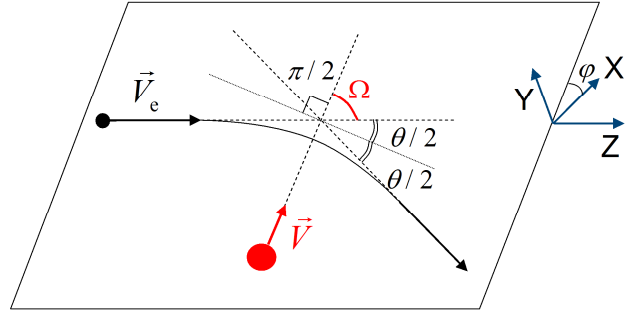

**Fig. 2.** Scheme of momentum transfer from an incident electron (black circle) to an atom nucleus (red circle). The velocity of the incident electron $\vec{V}_e$, velocity $\vec{V}$ of the atom after the collision, electron scattering angle $\theta$, atom emission angle $\Omega$ and azimuthal angle $\varphi$ are indicated.

For pure elastic collisions, the maximum energy transferred to a static atom corresponds to the scattering angle $\theta = \pi$ and is given by[66]

$$T_{max} = \frac{2 M E (E + 2 m_e c^2)}{(M + m_e)^2 c^2 + 2 M E}, \quad (6)$$

where $M$ is the atom mass, $E$ is the electron kinetic energy, while the angular dependence of the transferred energy is found as

$$T(\theta) = T_{max} \sin^2 \left( \frac{\theta}{2} \right). \quad (7)$$

In approach 1 for simulating electron collisions, the following algorithm is implemented. Minimal transferred energies $T_{min}[u]$ are assigned to each type $u$ of atoms. These minimal energies are chosen below the threshold energies for electron-induced structure transformations, such that they provide as full as possible a description of the reactions in the system at an affordable computational cost (typically minimal energies can be taken equal to 70 – 90% of threshold energies). The minimal angles for electron scattering are evaluated as

$$\theta_{min}[u] = 2 \arcsin \left( \sqrt{\frac{T_{min}[u]}{T_{max}}} \right). \quad (8)$$

One of the atoms $i$ is then selected randomly, based on the total cross-sections for each atom

$$\sigma_{tot}[u] = 2\pi \int_{\theta_{min}}^{\pi} \sigma(\theta) \sin \theta \, d\theta. \quad (9)$$

The angle $\theta$ of electron scattering is chosen in the interval $\theta \in [\theta_{min}[u_i]; \pi]$ according to the probability distribution $f(\theta) = 2\pi \sigma(\theta) \sin \theta / \sigma_{tot}[u_i]$. The emission angle of atom $i$ with respect to the electron incidence direction is found as $\Omega = (\pi - \theta)/2$ (Fig. 2). The azimuthal angle $\varphi$ is chosen in the interval $\varphi \in [0, 2\pi]$ with the uniform probability distribution.

Finally, the additional momentum $\Delta p = \sqrt{2MT(\theta)}$ is given to the selected atom $i$ at the chosen emission angle $\Omega$ and azimuthal angle $\varphi$.

In approach 2, principal electron-induced reactions are selected. It is assumed that these reactions take place for the impacted atom if the energy transfer exceeds some threshold energy for the reaction. In the simplest case, only reactions of atom emission are taken into account. The displacement threshold energy $E_d[u]$, i.e. the minimum kinetic energy required to remove the atom from its lattice position so that it does not immediately recombine, is assigned to each type $u$ of atoms. The total emission cross-section can be then evaluated as[67]

$$\sigma_d[u] = 4\pi \left(\frac{Ze^2}{2m_e c^2}\right)^2 \frac{1-\beta^2}{\beta^4} \left\{ \frac{T_{\max}}{E_d[u]} - 1 - \beta^2 \ln\left(\frac{T_{\max}}{E_d[u]}\right) \right. \\ \left. + \pi \frac{Ze^2}{\hbar c} \beta \left[ 2\left(\frac{T_{\max}}{E_d[u]}\right)^{1/2} - \ln\left(\frac{T_{\max}}{E_d[u]}\right) - 2 \right] \right\}. \quad (10)$$

Recently Meyer *et. al.*[68] demonstrated that taking into account the motion of the impacted atom due to lattice vibrations provides a closer match to experiment, introducing a smooth onset of knock-on damage. Therefore a more accurate expression for the maximum transferred energy can be used

$$T_{\max} = \frac{1}{2M} \left( r\left(r + \frac{2t}{c}\right) + \left(\frac{t}{c}\right)^2 \right), \quad (11)$$

$$r = \frac{1}{c}\sqrt{E(E + 2mc^2)} + MV_n, \quad (12)$$

$$t = \sqrt{(E + E_n)(E + 2mc^2 + E_n)}, \quad (13)$$

where $V_n$ is the initial velocity of the atom parallel to the electron beam and $E_n$ is the initial kinetic energy of the atom.

For the initial velocity $V_n$ of the atom parallel to the electron beam, the Gaussian probability distribution holds

$$P(V_n, T) dV_n = \frac{1}{\sqrt{\pi \langle v_T^2 \rangle}} \exp\left(-\frac{V_n^2}{\langle v_T^2 \rangle}\right) dV_n \quad (14)$$

The mean square velocity $\langle v_T^2 \rangle$ of the atom at temperature $T$ can be calculated according to the Debye model

$$\langle v_T^2 \rangle = \frac{9k_B}{8M}\theta_D + \frac{9k_BT}{M}\left(\frac{T}{\theta_D}\right)^3 \int_0^{\theta_D/T} \frac{x^3}{\exp(x)-1} dx, \quad (15)$$

where $k_B$ is the Boltzmann constant and $\theta_D$ is the Debye temperature. Using equations (11), (14) and (15) and integrating over all possible atomic velocities weighted by their probabilities, the maximum transferred energy and therefore the total displacement cross-section can be calculated with the thermal vibrations of the atom accounted for.

Alternatively, if the Debye temperature is unknown, the probability of the atomic velocities can simply be obtained directly from the Maxwell-Boltzmann velocity distribution

$$P(V_n, T) dV_n = \sqrt{\frac{M}{2\pi k_B T}} \exp\left(-\frac{MV_n^2}{2k_B T}\right) dV_n \quad (16)$$

This produces results that lie between those obtained using the static approach and those using the Debye temperature.

### 3.2 Classical MD simulations

The MD simulations of structure transformations under electron irradiation based on approach 1 include the following steps. First the system is equilibrated at temperature 300 K over 10 ps. The energy is then transferred to a random atom according to the algorithm described above, and the dissipation of this energy is modelled at a temperature of 300 K for 10 ps. Our simulations show that 10 ps is sufficient to capture all structural transformations induced by the electron collision. If no topological change is detected within this time period (the impact is unsuccessful), the simulation cycle is repeated. However, if the system topology has changed (the impact is successful), an additional simulation run is required to describe the structural relaxation between successive electron collisions. In experiments, successful electron collisions happen at intervals of several seconds, which is not accessible by atomistic methods. MD simulations of the motion of vacancies towards the edge of a graphene flake[58] imply that the average time necessary for the reconstruction of irradiation-induced defects in the flake is about 100 ps at a temperature of 2500 K. The estimation of the time $t_{th}$ necessary for the thermally induced graphene-fullerene transformation at 2500 K, using an approximation of the Arrhenius equation (3) based on MD simulations,[59] gives a reliable fulfilment of the condition (2). Molecular dynamics simulations of the processes that take place at these timescales are therefore performed at the elevated temperature of 2500 K over 100 ps. The simulations based on approach 2 with total cross-sections for atom emission are carried out in a similar way, but with an atom randomly chosen and removed with a probability determined by the total emission cross-sections.

The in-house MD-kMC[69] (Molecular Dynamics – kinetic Monte Carlo) code is used. The integration time step is 0.6 fs. The temperature is maintained by the Berendsen thermostat,[70] with a relaxation time of 0.1 ps at 300 K for equilibration of the system structure, 3 ps at 300 K for modelling electron collisions and 0.3 ps at 2500 K for modelling the structure relaxation between successful collisions. The covalent carbon-carbon interactions are described by the Brenner potential,[71] and the topology of the system is analysed on the basis of the "shortest-path" algorithm.[72] Two carbon atoms are considered to be bonded if the distance between them does not exceed 1.8 Å. Any atoms that detach completely from the system are removed to avoid reattachments.

### 3.3 Displacement threshold energy calculations

The displacement threshold energies, $E_d$, are calculated with AIMD simulations using the Q-Chem software package.[73] The structure of the nano-object is geometry optimised at the B3LYP/6-31G level, followed by a frequency calculation and a 1-step AIMD simulation at the same level of theory to generate the zero-point atomic velocities. For each type $u$ of atoms in the structure a representative atom is chosen. A series of AIMD simulations for each representative atom is run, in which the representative atom is assigned a range of velocities corresponding to potential values of $E_d$ and all other atoms are assigned zero-point atomic velocities. The range of velocities is chosen in order to find a minimum energy within 0.1 eV that permanently displaces the atom. As the interaction time of a relativistic electron with a nucleus is generally several orders of magnitude lower than the emission time of the atom, the total velocity is assigned at the beginning of the simulation.

A Fock matrix extrapolation procedure of using a 6$^{th}$-order polynomial and saving 12 Fock matrices was used in order to lower computational cost by using information from previous time steps to accelerate SCF convergence times. A time step of 1 fs was used, the SCF convergence criterion was $10^{-6}$ and the threshold for neglect of two electron integrals was $10^{-9}$.

## 4. Applications

### 4.1 Irradiation stability of terminating atoms in graphene nanoribbons

There are few prior AIMD simulations of single irradiation-induced events in systems containing atoms of multiple elements. These simulations have been restricted to studies relating only to the irradiation damage of nitrogen-doped graphene[9] and of h-BN.[52] This approach to modelling is applied here in relation to the formation of a new nanostructure under electron irradiation.

A sulphur-terminated graphene nanoribbon has been shown to self-assemble inside a SWNT from HRTEM electron irradiation of a functionalised fullerene.[25] Despite the presence of multiple hetero-elements (H, O, N and S) available for termination of the nanoribbon, sulphur termination was exclusively witnessed. This striking result can be explained by comparing displacement cross-sections of potential terminating atoms under electron irradiation, in this case hydrogen and sulphur.

The displacement threshold energies for the hydrogen, sulphur and carbon atoms terminating the nanoribbon edge have been calculated, as described in section 3.3, for the nanoribbon structure shown in Fig. 3. The same nanoribbon structure (including the terminating hydrogen atoms) was used for the simulations of the carbon atom displacement, as it has been shown that the presence of terminating hydrogen atoms has a negligible effect on the displacement threshold energy of edge carbon atoms in nanoribbons.[54] During the hydrogen displacement simulations the disruption to the structure of the nanoribbon was localised to a small area. However, during the sulphur and carbon atom displacement simulations the structure disruption extended to the ends of the short nanoribbon model. To more closely resemble the infinite nanoribbon without incurring unreasonably high computational expense, the carbon atoms at the two ends of the nanoribbon were fixed in position by assigning them arbitrarily high masses.

Displacement cross-sections have been calculated as described in section 3.1 for the considered hydrogen, sulphur and carbon atoms terminating the nanoribbon edge. The contributions of lattice vibrations were calculated using the Maxwell-Boltzmann velocity distribution, as the Debye temperature is unknown. The dependences of the calculated displacement cross-sections on the accelerating voltage are presented in Fig. 3. The calculated displacement threshold energies and displacement cross-sections at commonly used experimental conditions are listed in Table 1.

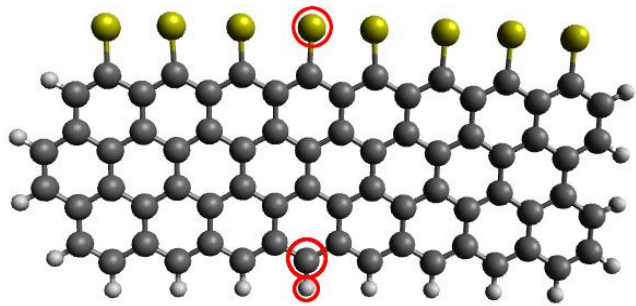

**Fig. 3.** Nanoribbon model used in the displacement threshold energy calculations. Carbon, hydrogen and sulphur atoms are shown by large dark grey, small light grey and yellow circles, respectively. Carbon, hydrogen and sulphur atoms chosen for the calculations are circled in red.

**Table 1.** Calculated displacement threshold energies $E_d$ (in eV) for terminating atoms of the graphene nanoribbon, and the corresponding displacement cross-sections $\sigma_d$ (in barn) at common experimental accelerating voltages.

| Atom | $E_d$ | $\sigma_d$ | |
|---|---|---|---|
| | | 80 kV | 200 kV |
| Hydrogen | 7.4 | 69.15 | 37.20 |
| Sulphur | 11.5 | 0 | 37.06 |
| Carbon | 16.5 | 0.53 | 23.91 |

At lower energies (<100 keV) the cross-section for hydrogen is extremely large relative to those of carbon and sulphur, rising to a peak of 355 barn at 7 keV. At an accelerating voltage of 80 kV, corresponding to experimental conditions,[25] it can be seen that while sulphur will undergo no emission from the nanoribbon, the displacement cross-section of hydrogen is very large. This difference in the sulphur and hydrogen atom displacement cross-sections is the reason for the preferential sulphur termination of the nanoribbon observed at the low accelerating voltages. At higher accelerating voltages however, sulphur (at 200 keV) and eventually carbon have larger cross-sections than hydrogen. At high electron kinetic energies, hydrogen will therefore counter-intuitively have the lowest probability of the three atoms of displacement from the nanoribbon edge. It can therefore be proposed that a hydrogen-terminated graphene nanoribbon can form instead of a sulphur-terminated nanoribbon, using the same precursor material but at a higher accelerating voltage. The possibility of different chemical reactions under electron irradiation in identical conditions other than electron kinetic energy means that electron irradiation can be used for selective control of chemical reactions.

The decrease in cross-sections at high accelerating voltages, seen most strikingly in the case of hydrogen atoms, can be understood by considering the length of the interaction time between the electron and the nucleus. As the electron kinetic energy increases, the time spent in close proximity to the atom decreases; $\beta$ (the electron velocity as a fraction of the speed of light) tends to one. As can be seen by equation (10), this reduces the cross-section, counteracting the increase in cross-section due to a larger transferred energy. At a certain accelerating voltage, the effect of the increase in $\beta$ outweighs the increase in the transferred energy, reducing the cross-section.

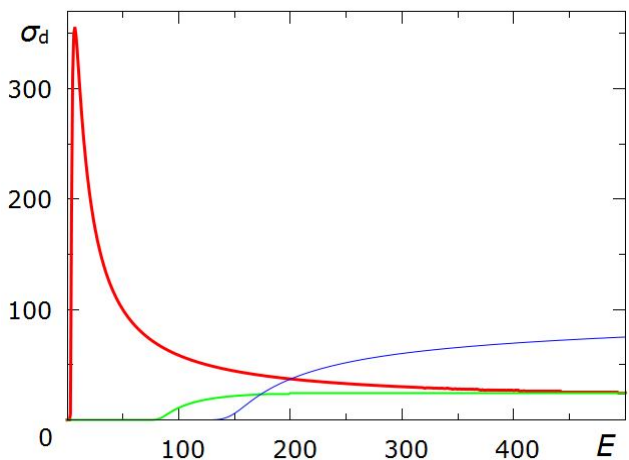

**Fig. 4.** Calculated displacement cross-sections $\sigma_d$ (in barn) as functions of electron kinetic energy $E$ (in keV) for hydrogen (thick red line), carbon (green line), and sulphur (thin blue line) atoms circled in **Fig. 3**.

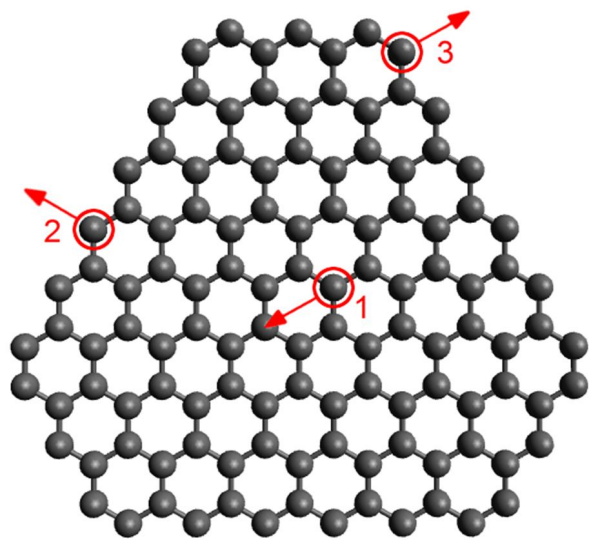

**Fig. 5.** Graphene flake consisting of 117 atoms considered in our simulations. Positions of atoms for which threshold energies were calculated are circled in red: (1) interior, (2) edge, (3) corner. Directions of momenta transferred in the flake plane for determination of threshold energies are indicated by red arrows.

Although the values for $E_d$ differ depending on the precise structure of the molecule, small changes in $E_d$ do not change the general trends shown in Fig. 4. In addition, the differences in cross-sections between atoms of different elements is generally much larger than the differences between atoms of the same element in different environments. This is because although different environments can alter $E_d$, atoms of different elements also have large differences in the maximum transmitted energy - as $Z$ increases, the energy transferred to the atom decreases. This means that while different molecules or nanostructures will have slight differences to the dependence of the cross-section on the accelerating voltage, Fig. 4 can be taken as a rough estimate of the general case for similar systems involving these elements.

The overlap of the carbon and hydrogen cross-sections has important implications for organic and biological TEM. In addition to the much higher levels of irradiation damage caused to these molecules by inelastic electron scattering processes such as ionisation and radiolysis,[74] Fig. 4 shows that there is no accelerating voltage at which predominantly carbon- and hydrogen-containing molecules will be stable under TEM irradiation.

### 4.2 Transformation of a graphene flake to fullerene

To compare approaches 1 and 2 to simulating electron irradiation, both have been applied to modelling a graphene-fullerene transformation. A graphene flake consisting of 117 atoms was considered (see Fig. 5) in the (20 nm x 20 nm x 20 nm) simulation cell. The kinetic energy of electrons in the beam was 80 keV and the electron flux was $4.1 \cdot 10^6$ electrons/(s·nm$^2$).[61,75] The relaxation of the flake between electron collisions and the equilibration of the flake before electron collisions were performed similarly in both approaches (as described in section 3.2). Only the difference in the description of electron collisions was studied.

In the simulations based on approach 2, only processes involving atom emission were taken into account, and the values of the cross-sections were determined previously.[61] To choose minimal transferred energies for the simulations based on approach 1, threshold energies for electron-induced structural reconstructions of the flake have been recalculated according to the semi-empirical potential[71] used. The threshold energies found for the emission of three-coordinated atoms in the flake interior, two-coordinated atoms at the flake edges and in chains at the flake corners (see Fig. 5) are listed in Table 2 and are in reasonable agreement with the displacement threshold energies obtained previously. For example, the calculated threshold energy for three-coordinated atom emission from the flake interior with the direction of transferred momentum perpendicular to the plane of the flake, 22.2 eV, is close to the values 22 eV[50] and 23 eV[47] obtained by DFTB calculations and 22.03 eV[52] obtained by AIMD calculations. It is worth noting that even these zero-temperature calculations of threshold energies reveal that some structural rearrangements can be induced in the graphene flake at lower transferred energies than atom emission. For example, an electron collision with a two-coordinated atom at the flake corner leads to the breaking of its bond with the neighbouring three-coordinated atom followed by the bond rearrangement for the neighbouring two-coordinated atom, starting from 13.9 eV and 18.3 eV in the cases of in-plane and out-of-plane momentum transfer, respectively. An electron collision with a two-coordinated atom at the flake edge already results in the formation of a ten-membered ring at 15.0 eV in the case of in-plane momentum transfer.

In addition to low-barrier reactions with no atom emission, it should be taken into account that threshold energies can decrease during the graphene-fullerene transformation due to the increase of curvature. Furthermore, at finite temperatures structural transformations in the flake can start at lower transferred energies than those determined at zero temperature. Thus minimal transferred energies smaller than the threshold energies for atom emission were chosen. For two-coordinated atoms at flake edges and corners, the minimal transferred energy was taken to be 10 eV. The same minimal energy was used for carbon atoms in defects (pentagons, heptagons and octagons), chains of carbon atoms, and three-coordinated atoms close to the flake edges (not farther than two bonds from the edge atoms). The minimum transferred energy for three-coordinated atoms in the flake interior was taken as 17 eV. The minimum transferred energy for one-coordinated carbon atoms was taken to be small (3 eV) to provide efficient elimination of such atoms. Our simulations show that these choices of minimal transferred energies result in 20% of the electron collisions that are considered leading to structural transformations of the graphene flake (they are successful).

**Table 2** Calculated threshold energies (in eV) for atom emission induced in the graphene flake by electron collisions at zero temperature for different directions of transferred momentum ("out-of-plane" corresponds to the direction perpendicular to the flake plane; "in-plane" corresponds to the direction parallel to the flake plane but perpendicular to the flake edge, see Fig. 5)

| Position of target atom | Momentum transfer | |
|---|---|---|
| | In-plane | Out-of-plane |
| 1. Interior | > 50 | 22.2 |
| 2. Edge | 16.3 | 20.9 |
| 3. Corner | 20.1 | 25.6 |

To detect the formation of a closed carbon cage the number of two-coordinated atoms, $N_2$, was monitored during the simulations. In the ideal flat flake this number should change with the total number $N$ of atoms in the flake as $N_2^0(N) \propto \sqrt{N}$. We assume that the fullerene formation occurs when the number of two-coordinated atoms in the carbon cluster, $N_2$, reaches half of the number of two-coordinated atoms in the ideal flat flake, $N_2^0(N)$ calculated for the total number $N$ of atoms at the given moment of time $\tau$, i.e. $N_2(\tau)/N_2^0(N(\tau)) = 1/2$. Correspondingly, this time $\tau$ is considered as the time of the graphene-fullerene transformation. The number of non-hexagonal polygons $N_d$ was also monitored for the analysis of structure evolution.

A comparison can now be made between the results obtained in 30 simulations on the basis of approach 1 and in 60 simulations on the basis of approach 2. Both of these approaches predict that the graphene flake can transform to a fullerene under electron irradiation (Fig. 6 and Fig. 7). The stages leading to fullerene formation are seen to be qualitatively similar for both approaches. As the atoms at the flake edges and corners have lower threshold energies, this is the location of most structural rearrangements. The very first steps of the simulations reveal the formation of non-hexagonal polygons at the flake edges (Fig. 6b and Fig. 7b). It can be also seen from the lower panels of Fig. 6l and Fig. 7n that the relative number of non-hexagonal polygons in the flake increases. Incorporation of pentagons at the edges leads to the formation of slightly curved flakes (Fig. 6c and Fig. 7c). These pentagons can be destroyed, leading to a flat structure (Fig. 6d), and can reappear (Fig. 6e). Fluctuations of the numbers of different polygons can be observed during this time (Fig. 6l). At some moment the reconstruction of the flake edges results in the formation of a bowl-shaped region (Fig. 6f and Fig. 7d), which can also disappear (Fig. 7e). A few attempts at the formation of such a bowl-shaped region followed by its destruction can be observed (Fig. 7f,g). As a result of these rearrangements, the shape of the flake can change dramatically (Fig. 7g) and the relative number of non-hexagonal polygons increases (Fig. 6l and Fig. 7n). The relative number of two-coordinated atoms $N_2/N_2^0$ however stays around unity. Finally, the formation of a bowl-shaped region (Fig. 6f and Fig. 7h) is followed by a fast zipping of the flake edges (Fig. 6g and Fig. 7i-k) and the formation of a closed carbon cage (Fig. 6h and Fig. 7l). In Fig. 6l and Fig. 7n this corresponds to a dramatic decrease in the relative number of two-coordinated atoms $N_2/N_2^0$ and a considerable increase in the number of non-hexagonal polygons. However, further structure relaxation favours the elimination of polygons with more than 6 vertices. An excess of pentagons also disappears (Fig. 6l and Fig. 7n). Despite the qualitatively similar behaviour of both approaches, the quantitative characteristics of the graphene-fullerene transformation and the further evolution of fullerene structure are quite different. First of all, the probability of observing the fullerene formation is relatively small (~40%) in the case of approach 2, and reaches 100% in the case of approach 1.

Predictions regarding the size of the fullerene formed also differ drastically. In the simulations based on approach 2, the distribution of sizes of fullerenes formed from the graphene flake has an average of 67 atoms and a root-mean-square deviation of 12 atoms. With further atom emission the fullerenes gradually decrease in size (Fig. 6i,j) until the curvature becomes too high to maintain the closure of the carbon cage and it opens. This takes place when the fullerenes contain about 50 atoms. Below this size, the carbon clusters evolve into structures less ordered than that of a flat flake (Fig. 6k), and at the last transformation stage weakly connected fragments of monoatomic linear chains and loops can be seen. It should be also noted that even during the stage of stable fullerene with more than 50 atoms, the number of non-hexagonal polygons fluctuates significantly (Fig. 6l) as electron collisions lead to the formation of new vacancies.

In the simulations based on approach 1, the loss of very few atoms is observed. The distribution of sizes of fullerenes formed from the graphene flake has an average of 115.9 atoms and a root-mean-square deviation of 1.4 atoms, i.e. on average only one atom is lost before the carbon cage closes (Fig. 7n). After the complete closure of the carbon cage (with less than 10 two-coordinated atoms left) no atom emission is observed at times accessible for our simulations (~ 10 minutes of real experimental time). The reactions taking place in the fullerene lead to the gradual elimination of polygons larger than hexagons and two-coordinated atoms (Fig. 7n), and finally a more ideal structure is formed (Fig. 7m).

Thus the majority of the transformations of the graphene flake induced by electron collisions should not lead to a loss of atoms.

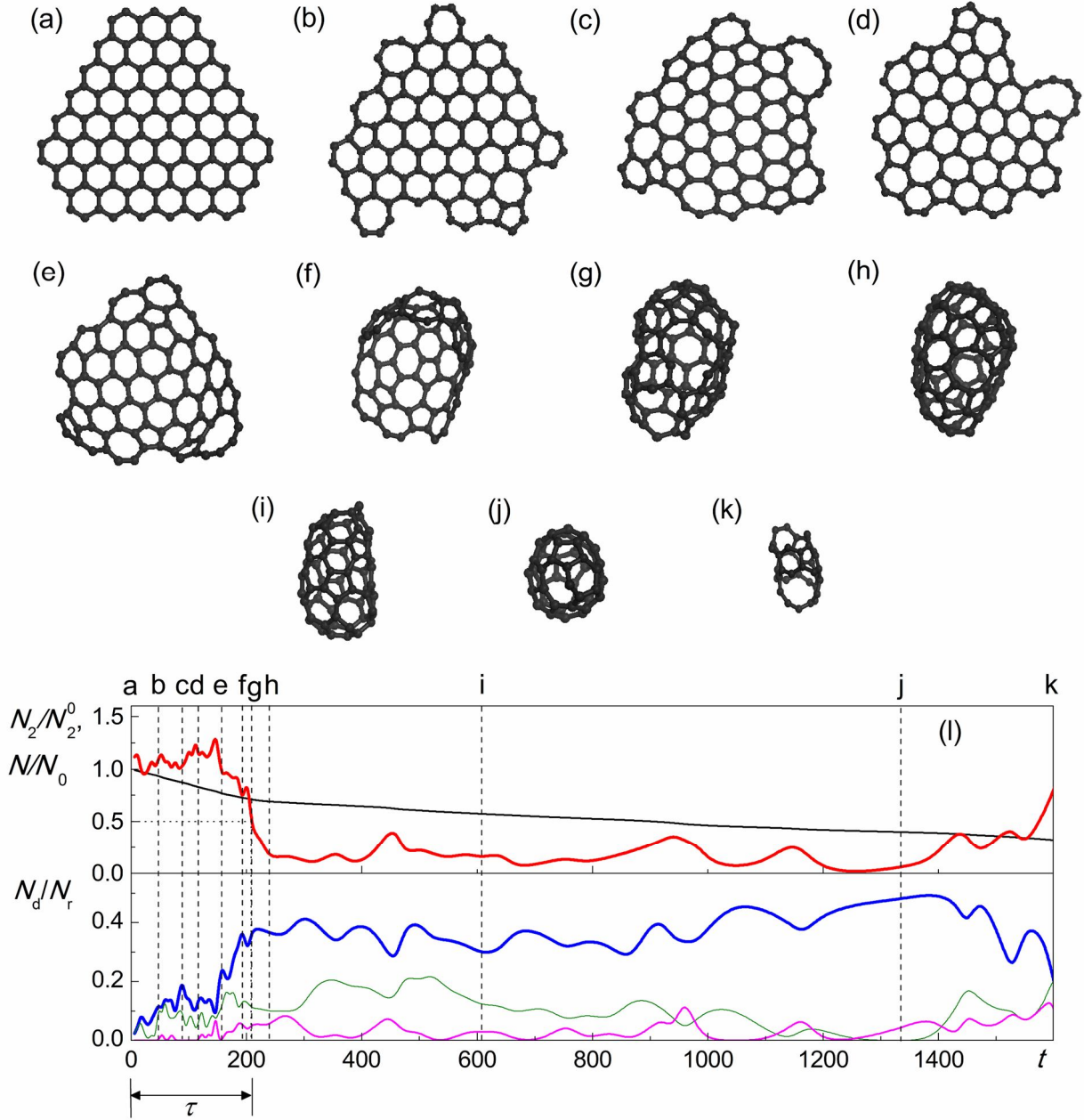

**Fig. 6.** (a–k) Evolution of structure of the graphene flake under irradiation by electrons with kinetic energy 80 keV and flux $4.1 \cdot 10^6$ electrons/(s·nm$^2$) observed in the simulations based on total cross-sections for atom emission: (a) 0 s, (b) 48 s, (c) 89 s, (d) 117 s, (e) 157 s, (f) 193 s, (g) 209 s, (h) 239 s, (i) 608 s, (j) 1335 s and (k) 1612 s. (l) Calculated number $N_2$ of two-coordinated atoms relative to the number $N_2^0$ of two-coordinated atoms in the ideal flake with the same total number $N$ of atoms (red line), total number $N$ of atoms in the flake relative to the initial value $N_0$ (black line), numbers $N_d$ of pentagons (thick blue line), heptagons (thin green line) and octagons (thin magenta line) relative to the total number $N_r$ of rings in the flake as functions of time $t$ (in s). Moments of time corresponding to structures (a–k) are shown using vertical dashed lines. The time $\tau$ of graphene-fullerene transformation is indicated by dotted lines and a double-headed arrow.

The average time between such reactions in the flat flake is found to be 2.0 s, which is considerably smaller than the interval of 6.5 s between atom emission events estimated using the total cross-sections. In the simulations based on approach 1 the interval between successive events of atom emission is actually much greater, on the order of hundreds of seconds. This difference in the time intervals between atom emission events can be attributed to the fact that formula (10) for the calculation of total cross-sections is derived under the assumption that the threshold energy is an isotropic function, which is false as seen in Table 2.

The characteristic times of graphene-fullerene transformation are also observed to be different in the simulations based on approaches 1 and 2. The distribution of times of graphene-fullerene transformation in the simulations based on approach 2 has an average of 350 s and a root-mean-square deviation of 90 s. Although the simulations based on approach 1 in the flake are induced at a faster rate, the formation of a fullerene in these simulations occurs at almost half the speed. The distribution of times of graphene-fullerene transformations has an average of 630 s and a root-mean-square deviation of 210 s.

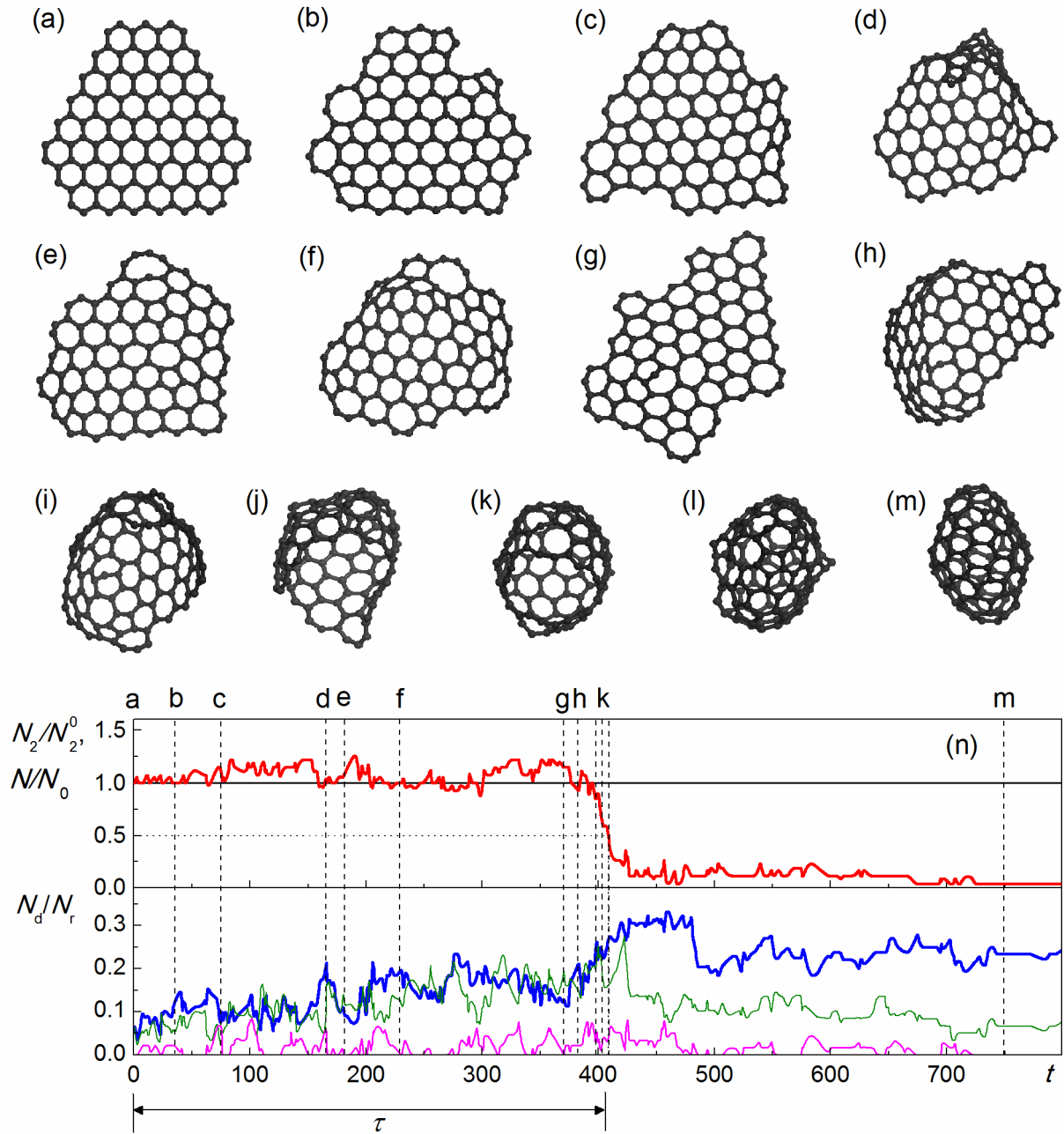

Fig. 7. (a–m) Evolution of structure of the graphene flake under irradiation by electrons with kinetic energy 80 keV and flux $4.1\cdot10^6$ electrons/(s·nm$^2$) observed in the simulations based on full description of energy and momentum transfer: (a) 0 s, (b) 36 s, (c) 76 s, (d) 166 s, (e) 181 s, (f) 230 s, (g) 370 s, (h) 384 s, (i) 398 s, (j) 401 s, (k) 408 s, (l) 411 s and (m) 750 s. (n) Calculated number $N_2$ of two-coordinated atoms relative to the number $N_2^0$ of two-coordinated atoms in the ideal flake with the same total number $N$ of atoms (red line), total number $N$ of atoms in the flake relative to the initial value $N_0$ (black line), numbers $N_d$ of pentagons (thick blue line), heptagons (thin green line) and octagons (thin magenta line) relative to the total number $N_r$ of rings in the flake as functions of time $t$ (in s). Moments of time corresponding to structures (a−m) are shown using vertical dashed lines. The time $\tau$ of graphene-fullerene transformation is indicated by dotted lines and a double-headed arrow.

Therefore it is clear that the full description of momentum and energy transfer represents a more rigorous approach for the simulation of electron irradiation. Nevertheless, this approach is much more computationally expensive than the use of total cross-sections for principal processes. The compromise between these methods can be achieved if the total cross-sections are calculated with an account of structure anisotropy, and if a more complete list of principal processes (beyond atom emission) is used.

## 5. Conclusions

Two approaches have been developed for the simulation of processes induced by electron irradiation: approach 1, which fully describes a transfer of energy and momentum, and approach 2, which uses total cross-sections for atom emission. The two approaches have been compared using the example of the transformation of a graphene flake to a fullerene. Both approaches predicted the same stages of the transformation: (1) formation of non-hexagonal polygons at flake edges, (2) transformation of the initially flat flake to a bowl-shaped structure, (3) fast zipping of flake edges and (4) further evolution of the fullerene under the action of the electron beam. However, the characteristics of the process were found to be rather different for approaches 1 and 2. The probability of observing the fullerene formation was 100% in the case of approach 1 and only 40% in the case of approach 2. No considerable loss of carbon atoms was detected in simulations of the basis of approach 1, contrary to approach 2. Correspondingly, the time between successive events of atom emission in the simulations based on approach 1 was considerably longer than in the simulations based on approach 2. Nevertheless, the characteristic time of fullerene formation was only twice as long for approach 1 as for approach 2. Thus the graphene-fullerene transformation is a bright example of a process induced by electron irradiation in which both atom emission and also reactions not resulting in atom loss play a significant role. Moreover, the difference in the intervals between atom emission events in the two approaches suggests that more complicated expressions taking into account structure anisotropy should be used for more accurate calculations of total cross-sections.

Each approach has inherent advantages, and is therefore applicable in different situations. In contrast to approach 2, approach 1 needs no prior knowledge of the exact values of cross-sections for possible irradiation-induced events for different types of atoms in the structure. A rough list of atom types and rough estimates of the threshold energy necessary for any irradiation-induced event to occur (for each type of atom) are sufficient. Thus approach 1 is easily useful even in the case where the local structure of the system cannot be predicted and analysed in advance, or where the number of atom types and/or the number of possible irradiation-induced events for each atom type is very large (for example in nanostructures containing atoms of several different elements). Moreover, in many cases of electron irradiation-assisted nanostructure transformation, the further behaviour of an atom subjected to an electron impact and removed from the structure has an essential influence on the transformation process. Key examples of this include processes inside carbon nanotubes and the irradiation damage of few-layer graphene. In these cases the immediate transfer of kinetic energy from the electron should be included in the model, meaning approach 1 should be used to simulate such irradiation-induced processes.

Approach 2 has the key advantage of being based on accurate values of cross-sections of irradiation-induced events. As approach 1 considers irradiation-induced events with the use of classical potentials for the description of interatomic interactions, it cannot currently ensure accurate values of these cross-sections. Thus, approach 2 combines the accuracy of *ab initio* calculations when considering a single irradiation-induced event with the time scale and large number of atoms of classical MD simulations when considering the whole irradiation-induced process. Additionally, experimental values of cross-sections can be used in approach 2. Currently, such values are only measured for some types of atoms in carbon networks,[68] which is insufficient when considering most irradiation-induced processes. However, it is likely that accurate values of cross-sections for various types of atoms in carbon nanostructures will be measured in the near future, as further progress is made in HRTEM studies.

As discussed in Section 2.2, the values of cross-sections of irradiation-induced events can depend considerably on the angle between the electron beam direction and the chemical bonds. To adequately use approach 2 it is therefore necessary not only to determine all possible types of atoms with different local structure, but also to take into account the dependence of the cross-sections on the angle between the electron beam direction and the chemical bonds. The application of approach 2 is therefore mostly confined to simulating irradiation-induced processes in flat 2D systems where the electron beam is perpendicular to the structure, in which this angle is equal for all atoms of the system. Examples of such processes include irradiation damage of suspended graphene monolayers or nanoribbons.

Theoretical modelling of processes initiated or assisted by electron irradiation can provide a detailed understanding of HRTEM images. Both approaches outlined in this article can be used to assign or confirm experimental observations, with approach 1 in particular providing possible predictive power. This can be used to aid the selection of optimal experimental conditions, for example to minimise damage to a structure or to focus the irradiation damage in a specific way. The example of the differing irradiation stabilities of terminating atoms in graphene nanoribbons illustrates this point, suggesting a possible way forward for tailoring selectivity in chemical reactions under the electron beam.

## Acknowledgements

EB acknowledges EPSRC Career Acceleration Fellowship, New Directions for EPSRC Research Leaders Award (EP/G005060), and ERC Starting Grant for financial support. AP acknowledges Russian Foundation of Basic Research (1202-90041-Bel and 11-02-00604-a) and Samsung Global Research Outreach Program. IL acknowledges support from Spanish grant (FIS2010-21282-C02-01), Grupos Consolidados del Gobierno Vasco (IT-319-07), and computational time on the Multipurpose Computing Complex NRC "Kurchatov Institute".

## Notes and references

*Corresponding author: Elena Bichoutskaia*
*E-mail: elena.bichoutskaia@nottingham.ac.uk*
[a] *Department of Chemistry, University of Nottingham, University Park, Nottingham NG7 2RD, United Kingdom.*
[b] *Kintech Lab Ltd., Kurchatov Square 1, Moscow 123182, Russia.*

*[c] Nano-Bio Spectroscopy Group and ETSF Scientific Development Centre, Departamento de Fisica de Materiales, Universidad del Pais Vasco UPV/EHU, San Sebastian E-20018, Spain.*
*[d] Institute of Spectroscopy of Russian Academy of Sciences, Fizicheskaya Street 5, Troitsk, Moscow 142190, Russia.*